\documentclass[10pt]{article}
\usepackage{graphicx}
\usepackage{xcolor}
\usepackage{amssymb}
\usepackage{amsmath}
\usepackage{nicefrac}
\usepackage{caption}
\usepackage{fullpage}

\newcommand{\equaref}[1]{Eq.~(\ref{#1})}

\newcommand{\figref}[1]{Fig.~\ref{#1}}

\newcommand{\secref}[1]{Section~\ref{#1}}

\usepackage[english]{babel}
\usepackage[utf8x]{inputenc}
\usepackage{amsmath}
\usepackage{graphicx}
\usepackage[colorinlistoftodos]{todonotes}
\usepackage{slashed}

\usepackage{fullpage}
\usepackage{caption}
\usepackage{subcaption}
\usepackage{amssymb}
\usepackage{color}
\usepackage[colorlinks = true,linkcolor = blue]{hyperref} 
\newcommand\pubnumber{}
\newcommand\pubdate{\today}

\def\imperialpd{Institute for Particle Physics Phenomenology, University of Durham, 
South Rd, DURHAM,
 DH13LE, UNITED KINGDOM}
\def\Title#1{\begin{center} {\Large #1 } \end{center}}

\def\Address#1{\begin{center}{ \it #1} \end{center}}

\newcommand\pubblock{\rightline{\begin{tabular}{l} \pubnumber\\
         \pubdate  \end{tabular}}}
\newenvironment{Abstract}{\begin{quotation}  }{\end{quotation}}
\newenvironment{Presented}{\begin{quotation} \begin{center}
             PRESENTED AT\end{center}\bigskip
      \begin{center}\begin{large}}{\end{large}\end{center} \end{quotation}}

\begin{document}
\begin{titlepage}
\pubblock

\vfill
\Title{Creating the Baryon Asymmetry from Lepto-Bubbles}
\vfill
\center{Silvia Pascoli, \textbf{Jessica Turner}, Ye-Ling Zhou}
\Address{\imperialpd}
\vfill
\begin{Abstract}
We propose a new mechanism of baryogenesis which proceeds via a CP-violating phase transition. During this phase transition, the coupling of the Weinberg operator is dynamically realised and subsequently a lepton asymmetry is generated via the non-zero interference of this operator at different times. This new scenario of leptogenesis provides a  direct connection between the baryon asymmetry, low energy neutrino parameters and leptonic flavour models. 
\end{Abstract}
\vfill
\begin{Presented}
NuPhys2016, Prospects in Neutrino Physics\\
Barbican Centre, London, UK,  December 12--14, 2016
\end{Presented}
\vfill
\end{titlepage}
\def\thefootnote{\fnsymbol{footnote}}
\setcounter{footnote}{0}

\newcommand{\startcompact}[1]{\par\vspace{-0.75em}\begin{#1}%
\allowdisplaybreaks\ignorespaces}

\newcommand{\stopcompact}[1]{\end{#1}\ignorespaces}

\section{Introduction}
In spite of the abundance of baryogenesis models, the  matter anti-matter asymmetry  remains an unresolved issue.  Common to all dynamical mechanisms of baryogenesis is the necessity to satisfy Sakharov's  conditions \cite{Sakharov:1967dj}: processes must violate baryon number, C and CP-symmetry and finally provide a departure from thermal equilibrium. Popular scenarios of high-scale thermal leptogenesis use the out-of-equilibrium decays of heavy particles, such as sterile neutrinos, to produce a lepton asymmetry which is subsequently converted to a baryon asymmetry via electroweak sphaleron processes \cite{Fukugita:1986hr}. In addition to providing a possible solution to the baryon asymmetry, thermal leptogenesis also explains small neutrino masses. 
There are also low-scale leptogenesis mechanisms  in which the mass of the sterile neutrinos is below the electroweak scale. The lepton asymmetry is generated via the oscillations of these sterile states \cite{Akhmedov:1998qx}. In both  high and low-scale leptogenesis, the details of the neutrino mass model must be specified.\\
In this paper we propose a completely new scenario of leptogenesis which proceeds via a CP-violating phase transition and henceforth will be referred to as the CPPT mechanism. This phase transition   produces bubbles which are leptonically CP-violating, namely  \emph{lepto-bubbles}. The generation of the lepton asymmetry occurs below the neutrino mass generation scale and therefore the particular neutrino mass model need not be specified. In addition to providing a novel solution to the matter anti-matter asymmetry, our mechanism establishes a connection between this asymmetry and the flavour structure of the lepton sector \cite{Pascoli:2016tiv, Pascoli:2016gkf}. \\
This short paper is structured as follows: in \secref{sec:mech} and \secref{sec:CTP} we discuss our basic assumptions, outline the CPPT mechanism and the computational method namely the Closed-Time Path Formalism. In \secref{sec:calc} we give details of the lepton asymmetry calculation and estimate the temperature at which the phase transition occurs and finally we discuss and summarise in \secref{sec:discussion} and \secref{sec:conclusion} respectively.

\section{The CPPT Mechanism}\label{sec:mech}
We assume neutrinos are Majorana in nature and therefore, to leading order, their mass model reduces to the lepton number violating Weinberg operator 
\startcompact{small}
\[\label{eq:Weinberg}
\mathcal{L}_{W}=\frac{\lambda_{\alpha\beta}}{\Lambda}\ell_{\alpha L}H C\ell_{\beta L} H + \text{h.c.} \,,
\]
\stopcompact{small}
where $\lambda_{\alpha\beta}=\lambda_{\beta\alpha}$ is a model-dependent coupling,  $\Lambda$  the scale of new physics  and  $C$ is the charge conjugation matrix. The Weinberg operator can be UV-completed in a number of ways ranging from  loop effects to introducing heavy new degrees of freedom such as sterile neutrinos. However, unlike typical scenarios of leptogenesis, the details of the UV-completion of the dimension five operator need not be specified in this mechanism. We postulate the coupling of the Weinberg operator is functionally dependent upon a SM-singlet scalar, $\phi$, such that $\lambda_{\alpha\beta}=\lambda^{0}_{\alpha\beta}+\lambda^{1}_{\alpha\beta}\langle\phi\rangle/v_{\phi}$.  Associated to $\phi$ is a finite temperature scalar potential, which is symmetric under a leptonic flavour symmetry at sufficiently high temperatures.  As the temperature of the Universe lowers, the minima at the origin of this potential becomes metastable and a phase transition occurs. As a result, the minima changes from the vacua at the origin to a deeper, true vacua which is stable and non-zero, $\langle \phi\rangle$.  The ensemble expectation value (EEV) of $\phi$ spontaneously breaks the high-scale flavour symmetry and results in the  observed pattern of  leptonic masses and mixing. Assuming a first order phase transition, (lepto) bubbles of the leptonically CP-violating broken phase spontaneously nucleate. At a fixed space point within the bubble wall, the coupling $\lambda$ is time-dependent e.g.  $\lambda\left(t_1\right) \neq \lambda\left(t_2\right)$ for $t_{1}\neq t_{2}$. As a consequence,  the lepton asymmetry arises from the  non-zero interference of the Weinberg operator at different times.

\section{The Closed-Time Path Formalism}\label{sec:CTP}
We  apply the closed-time path (CTP) formalism to calculate the lepton asymmetry. Unlike zero temperature methods, the CTP formalism properly accounts for finite density effects and  resolves unitarity issues. The basic building blocks of the CTP formalism are the Green functions and the corresponding self-energy corrections.  As we will focus on the self-energy corrections from the Weinberg operator to the lepton propagators, it is sufficient to present the  Higgs ($\Delta$) and lepton ($S$)  propagators
\startcompact{small}
\begin{eqnarray*}
 \Delta^{T,\overline{T}}(x_1,x_2) &=& \langle T[H(x_1) H^*(x_2)] \rangle, \langle \overline{T}[H(x_1) H^*(x_2)] \rangle \,,\\
 \Delta^{<,>}(x_1,x_2) &=& ~~ \langle H^*(x_2) H(x_1) \rangle,~~~~\langle H(x_1) H^*(x_2) \rangle \,, \\
 S^{T,\overline{T}}_{\alpha\beta}(x_1,x_2) &=& \langle T[\ell_\alpha(x_1) \overline{\ell}_\beta(x_2)] \rangle, \langle \overline{T}[\ell_\alpha(x_1) \overline{\ell}_\beta(x_2)] \rangle \,, 
\\
 S^{<,>}_{\alpha\beta}(x_1,x_2) &=& -\langle \overline{\ell}_\beta(x_2) \ell_\alpha(x_1) \rangle , ~~~~\langle \ell_\alpha(x_1) \overline{\ell}_\beta(x_2) \rangle \,,
\label{eq:fermion_propagator}
\end{eqnarray*}
\stopcompact{small}
where $T$ ($\overline{T}$) denotes time (anti-time) ordering, Greek indices are flavour indices and spinor and electroweak indices have been suppressed. The lepton asymmetry is written in terms of the leptonic Wightman propagators, $S^{<,>}$, such that
\startcompact{small}
\begin{equation*}
 n_L(x)  =  - \frac{1}{2} \sum_\alpha \text{tr}\Big\{  \gamma^0 \big[S^<_{\alpha\alpha}(x,x) + S^>_{\alpha\alpha}(x,x) \big] \Big\} .
\label{eq:current}
\end{equation*}
\stopcompact{small}
The Kadanoff-Baym  (KB) equations are used to calculate the time evolution of the lepton asymmetry and we follow the conventions of \cite{Prokopec:2003pj,Garbrecht:2008cb} and write this equation as
\startcompact{small}
\begin{equation}
i \slashed{\partial}  S^{<,>} - \Sigma^{H} \odot S^{<,>} - \Sigma^{<,>} \odot S^{H} 
 = \frac{1}{2} \big[\Sigma^{>} \odot S^{<} - \Sigma^{<} \odot S^{>} \big] \,,
\label{eq:KB}
\end{equation}
\stopcompact{small}
where the symbol $\odot$ presents a convolution, $\Sigma$ is the self-energy of the lepton and $S^H$ ($\Sigma^H$) is Hermitian parts of propagator (self-energy) given by $S^H=S^T-\frac{1}{2}\left(S^>+S^<\right)$ ($\Sigma^H=\Sigma^T-\frac{1}{2}\left(\Sigma^>+\Sigma^<\right)$). The self-energy contribution ($\Sigma^{H}S^{<,>}$) and broadening of the on-shell dispersion relation ($\Sigma^{<,>} S^{H}$) are given on the LHS of \equaref{eq:KB}. Whilst, the collision term that includes  CP-violating source ($\frac{1}{2} (\Sigma^{>} S^{<} - \Sigma^{<} S^{>})$) is shown on the RHS \cite{Prokopec:2003pj}.  As we  focus on the generation of an initial asymmetry, we consider only the collision term.
 
\startcompact{small}
\begin{figure}[t]
\centering
\begin{minipage}{.4\textwidth}
  \centering
  \includegraphics[width=1.0\linewidth]{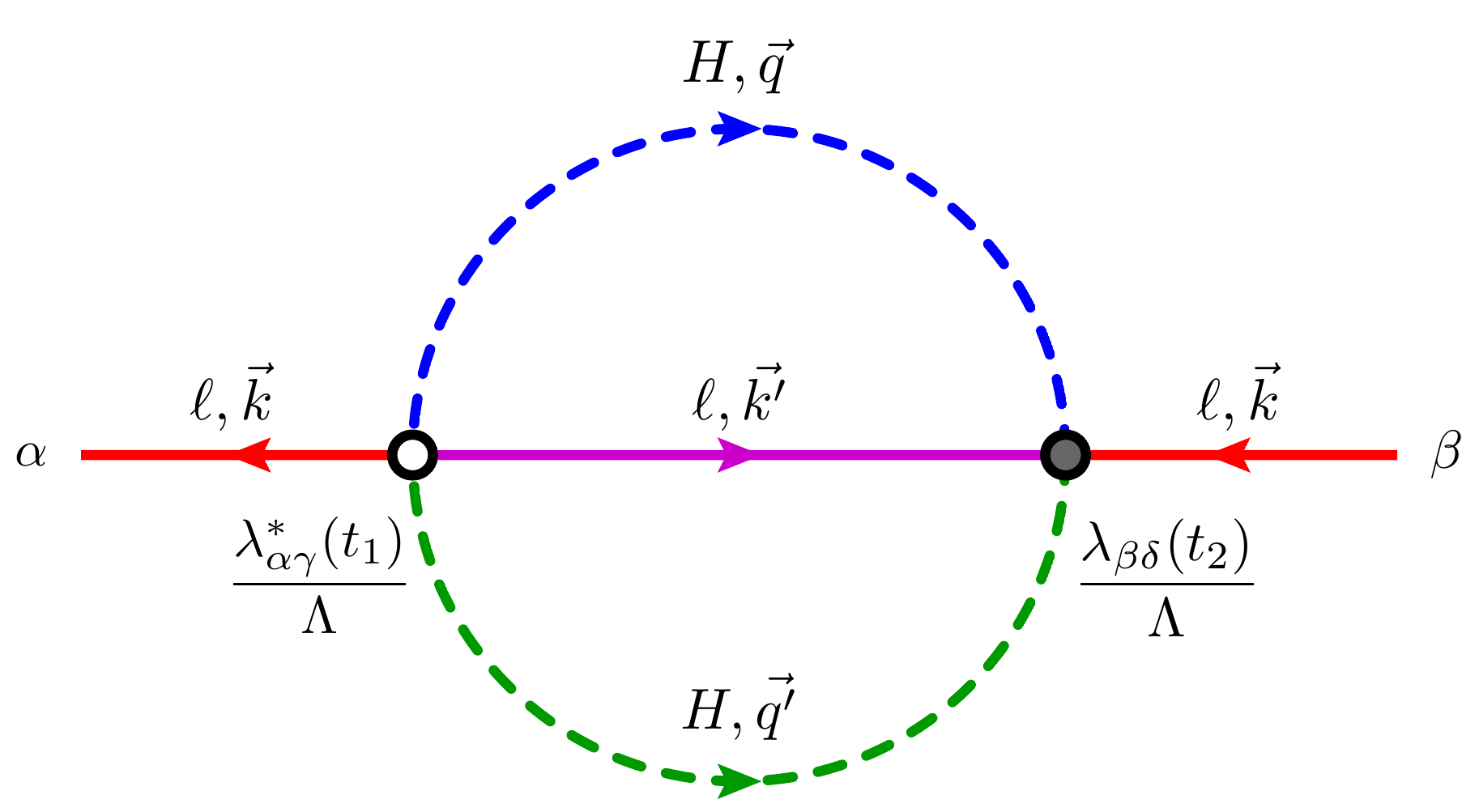}
  \vspace{0.2cm}
  \captionof{figure}{Two loop lepton number violating contribution of the CP-violating and time dependent Weinberg operator to the lepton self-energy.}
  \label{fig:feyn}
\end{minipage}%
\hfill
\begin{minipage}{.4\textwidth}
  \centering
  \includegraphics[width=1.0\linewidth]{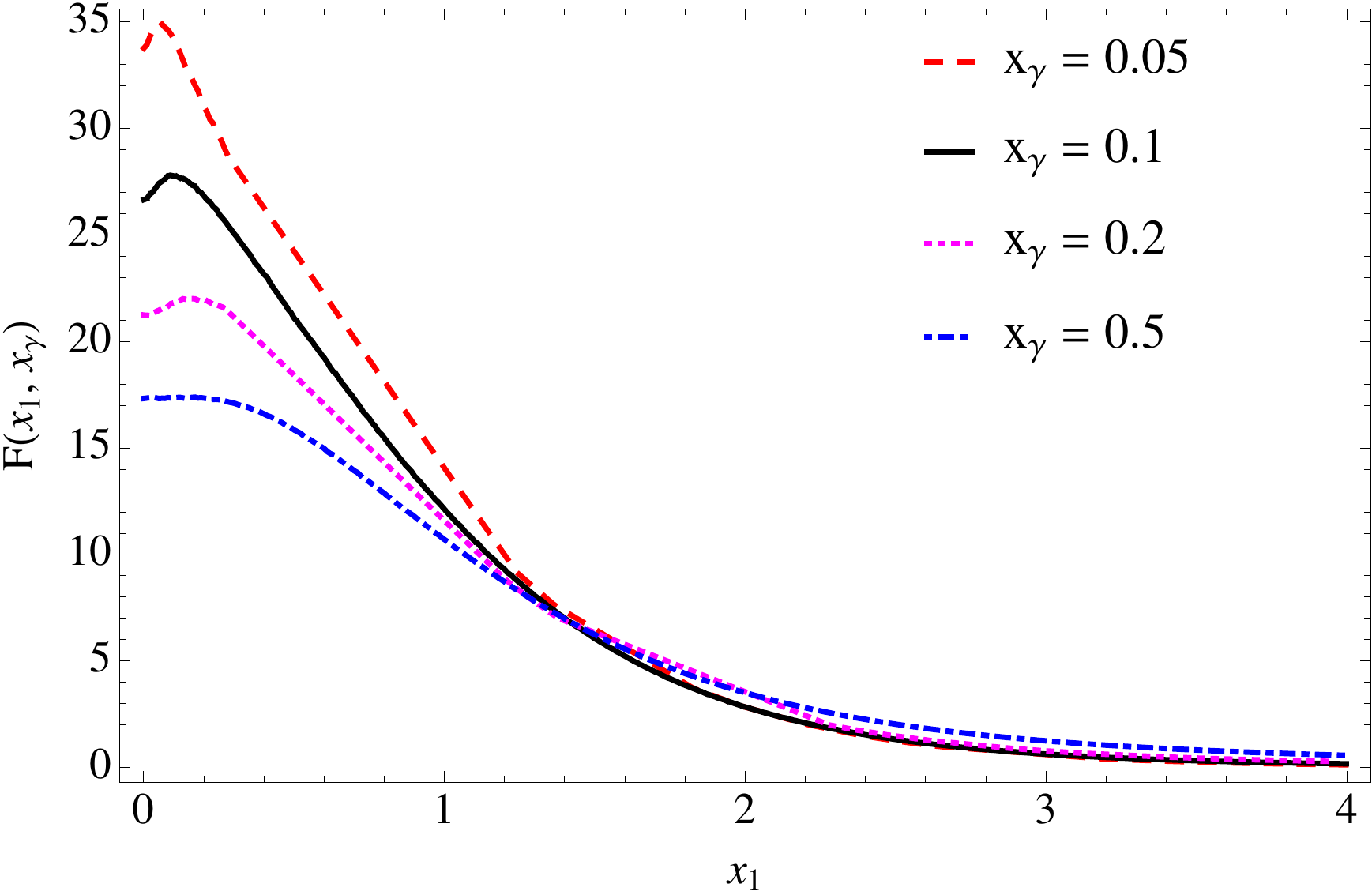}
  \captionof{figure}{The loop factor $F(x_1,x_\gamma)$ as a function of the lepton energy $k$ and the thermal width $\gamma$, where $x_1 = k/2T$ and $x_\gamma = \gamma/T$.}
  \label{fig:loopfactor}
\end{minipage}
\end{figure}
\stopcompact{small}

\section{Calculating the Lepton Asymmetry}\label{sec:calc}
We calculate the lepton asymmetry using similar techniques applied in \cite{Anisimov:2010dk}. The calculation involves application of the KB equation \eqref{eq:KB} to find the time evolution of the lepton propagator.  The self-energy correction is calculated to leading order, in the time-independent flavour basis,  as is shown in \figref{fig:feyn}. The first step is to Fourier transform the Green functions of the Higgs and leptonic propagator respectively
  
\startcompact{small}
\begin{equation*}
\Delta_{\vec{q}}(t_1,t_2) = \int d^3 r e^{i \vec{q} \cdot \vec{r}} \Delta(x_1,x_2) \quad \text{and} \quad
S_{\vec{k}}(t_1,t_2) = \int d^3 r e^{i \vec{k} \cdot \vec{r}} S(x_1,x_2)\,,
\end{equation*}
\stopcompact{small}
where $t_{1}=x_{1}^0$, $t_{2}=x_{2}^0$ and $r=x_{1}-x_{2}$.  The lepton asymmetry, at a fixed space point in the bubble wall,  is given by $n_L(x) = \int \frac{d^3 k}{(2\pi)^2} L_{\vec{k}}$ with
\startcompact{small}
\begin{equation} 
\begin{aligned} 
L_{\vec{k}} \equiv f_{\ell\vec{k}} - f_{\overline{\ell}\vec{k}} 
&= - \int_{t_i}^{t_f} dt_1 \partial_{t_1} \text{tr}[\gamma_0 S^<_{\vec{k}}(t_1,t_1) + \gamma_0 S^>_{\vec{k}}(t_1,t_1)] \\
	 &=  -\int_{t_{i}}^{t_{f}} dt_{1}\int_{t_{i}}^{t_{f}} dt_{2} \text{tr} \Big[\Sigma^{>}_{\vec{k}}(t_{1},t_{2})S_{\vec{k}}^{<}(t_{2},t_{1})  
  -\Sigma^{<}_{\vec{k}}(t_{1},t_{2})S_{\vec{k}}^{>}(t_{2},t_{1})  \Big] \,,
 \label{eq:key}
\end{aligned}
\end{equation}
\stopcompact{small}
where $t_{i}$ ($t_{f}$) is the initial (final) time and $\Sigma^{<}_{\vec{k}}(t_{1},t_{2})$ is the   CP-violating two-loop self-energy correction as shown in \figref{fig:feyn}. Subsequently, \eqref{eq:key} may be re-expressed as
\startcompact{small}
\begin{equation*}
L_{\vec{k}\alpha\beta} = \sum_{\gamma\delta}\frac{12}{\Lambda^{2}}\int_{t_{i}}^{t_{f}} dt_{1}\int_{t_{i}}^{t_{f}} dt_{2} \text{Im}\big\{ \lambda^{*}_{\alpha\gamma}(t_{1})\lambda_{\beta						\delta}(t_{2})\big\} 
					\int_{q, q^{\prime}} M_{\alpha\beta\gamma\delta}(t_1,t_2,k,k^{\prime},q,q^{\prime}) \,,
\end{equation*}
\stopcompact{small}
where $\int_{q, q^{\prime}}=\int \frac{d^{3}q}{(2\pi)^3} \frac{d^{3}q^{\prime}}{(2\pi)^3}$ . The lepton asymmetry has factorised into two parts:  
one part is a function of the time-dependent couplings, $\lambda\left(t\right)$, which allows for a connection between the lepton asymmetry, the leptonic flavour model and low energy neutrino parameters. The other, $M_{\alpha\beta\gamma\delta}(t_1,t_2,k,k^{\prime},q,q^{\prime})$,  is the finite temperature matrix element which is calculated using CTP Feynman rules (for further details see \cite{Anisimov:2010dk}). For the present calculation, we will ignore the differing thermal widths of the charged lepton propagators and using the limit
 $t_i (t_{f})\to -\infty (+\infty)$, the total lepton asymmetry $L_{\vec{k}} \equiv \sum_{\alpha} L_{\vec{k}\alpha\alpha}$ is given by 
 \startcompact{small}
\begin{eqnarray*}
\hspace{-5mm}L_{\vec{k}} = \frac{12}{\Lambda^{2}} \int_{-\infty}^{+\infty} \!\!\!\!dt_{1}\int_{-\infty}^{+\infty} \!\!\!\!dt_{2}  \text{Im}\big\{ \text{tr} \left[ \lambda^{*}(t_{1})\lambda(t_{2})\right]\big\} \int_{q, q^{\prime}} M \,,
\end{eqnarray*}
\stopcompact{small}
where the  finite temperature matrix element, decomposed in terms of the lepton and Higgs propagators, is expressed as
\startcompact{small}
\begin{eqnarray*}\label{eq:ME}
M = \text{Im}\big\{ \Delta^{<}_{\vec{q}}(t_1,t_2)  \Delta^{<}_{\vec{q^{\prime}}}(t_1,t_2) \text{tr} \big[ S^{<}_{\vec{k}}(t_1,t_2)S^{<}_{\vec{k^{\prime}}}(t_1,t_2) P_{L}  \big]   \big\}\,. 
\end{eqnarray*}
\stopcompact{small}
Throughout we assume the  Higgs and leptonic propagators are almost in thermal equilibrium as the scale of the CPPT mechanism is significantly higher than the electroweak scale.
The time-varying coupling is functionally dependent upon the EEV of the scalar, $\phi$.  We make an ansatz for the coupling
\startcompact{small}
\begin{equation*}
\lambda(t) =\lambda^0+ \lambda^1 f(t),
\end{equation*}  
\stopcompact{small}
where  $\lambda^0$ ($\lambda^0+\lambda^1$) is the value of the coupling at $t=-\infty$ ($t=+\infty$) and $f(t)$  varies continuously from $0$ to $1$.
As expected, the lepton asymmetry is not sensitive to the  precise functional form of $f(t)$; it has been shown a $\tanh$ and step function produce the same result \cite{Pascoli:2016gkf}.  Performing a change of integration variables from $t_{1}$, $t_{2}$ to  $\tilde{t}=\left(t_{1}+t_{2}\right)/2$,  $y=t_{1}-t_{2}$ and using $\int_{-\infty}^{+\infty} d\tilde{t} [f(\tilde{t}+y/2)-f(\tilde{t}-y/2)] = y$, the lepton asymmetry may be written as 
\startcompact{small}
\begin{equation}
\label{eq:asymmetry}
L_{\vec{k}} 
=  - \frac{12}{v^{4}_{H}} \text{Im}\{ \text{tr}[ m^{0}_{\nu}m_{\nu}^* ]\} \int_{-\infty}^{+\infty}dy y \int_{q,q^{\prime}} M \,,
\end{equation}
\stopcompact{small}
where $v_{H}$ is the Higgs vacuum expectation value and $m_\nu^0 \equiv \lambda^0 v_H^2/\Lambda$ $\left(m_\nu  \equiv (\lambda^0 + \lambda^1) v_H^2/\Lambda\right)$ is the effective neutrino mass matrix before (after) the phase transition.  
As $\phi$ only interacts with the leptons and Higgs in the thermal bath, it is reasonable to assume a fast-moving bubble wall. Consequently, the lepton asymmetry \eqref{eq:asymmetry} is not dependent upon the bubble  properties. Using the calculation of  $M$ (full details given in \cite{Pascoli:2016gkf}) the lepton asymmetry can be rewritten as 
\startcompact{small}
\begin{equation*}
L_{\vec{k}} = \frac{3\,\text{Im}\Big\{ \text{tr}\left[ m^{0}_{\nu}m^{*}_{\nu} \right]  \Big\}T^2}{\left( 2\pi \right)^{4}v^{4}_{H}}  F\left( x_{1},x_{\gamma} \right)\,.
\end{equation*}
\stopcompact{small}
 $F\left( x_{1},x_{\gamma} \right)$ is a loop factor given by
 \startcompact{small}
\begin{eqnarray*}
F\left( x_{1},x_{\gamma} \right) = &&\frac{1}{x_{1}} \int_{0}^{+\infty}\!\!\!\!dx  \int_{0}^{+\infty}\!\!\!\!\!x_{2}dx_{2} \int_{\left|x_{1}-x \right|}^{\left|x_{1}+x \right|}\!\!\!\!dx_{3}  \int_{\left|x_{2}-x \right|}^{\left|x_{2}+x \right|}\!\!\!\!dx_{4}  \sum_{\eta_{2},\eta_{3},\eta_{4}=\pm1} \bigg[1-\frac{\left(x^2_{1}+x^{2}-x^{2}_3\right)\left(x^2_{2}+x^{2}-x^{2}_4\right)}{4\eta_{2}x_{1}x_{2}x^2} \bigg]\nonumber\\
&&\times \frac{X_{\eta_{2}\eta_{3}\eta_{4}}x_{\gamma}\sinh X_{\eta_{2}\eta_{3}\eta_{4}}}{\left(X^2_{\eta_{2}\eta_{3}\eta_{4}}+x^2_{\gamma}\right)^{2}\cosh x_{1}\cosh x_{2}\sinh x_{3}\sinh x_{4}} \,,
\end{eqnarray*}
\stopcompact{small}
where the  four momentum of the leptons and Higgs shown in \figref{fig:feyn} are defined as $k = |\vec{k}|$, $k = |\vec{k^{\prime}}|$,  $q = |\vec{q}|$ and  $q = |\vec{q^{\prime}}|$ and correspondingly  $x_{1}=k/2T$,  $x_{2}=k^{\prime}/2T$,  $x_{3}=q/2T$,  $x_{4}=q^{\prime}/2T$,  $x=p/2T$ and  $X_{\eta_{2}\eta_{3}\eta_{4}}=x_{1}+\eta_{2}x_{2}+\eta_{3}x_{3}+\eta_{4}x_{4}$. The loop factor is dependent upon the lepton energy and the thermal width normalised by the temperature, i.e., $x_1$ and $x_\gamma$ as shown in \figref{fig:loopfactor}. 

\section{Discussion}\label{sec:discussion}

The lepton asymmetry produced during the CPPT mechanism is partially converted into a final baryon asymmetry via sphaleron processes which are unsuppressed above the electroweak scale. The final baryon asymmetry is roughly given by $\eta_{B}\approx\frac{1}{3}\eta_{B-L}$ and may be written as
\startcompact{small}
\begin{equation*}\label{eq:asym}
\frac{n_{B}}{n_{\gamma}} \approx -\frac{\,\text{Im}\Big\{ \text{tr}\left[ m^{0}_{\nu}m^{*}_{\nu} \right]  \Big\}T^2}{8\pi^2\zeta\left(3\right)v_{H}^4}  F\left( x_{\gamma} \right), \,
\end{equation*}
\stopcompact{small}
where  $F\left( x_{\gamma} \right)=\int_{0}^{\infty}x_{1}dx_{1}F\left( x_{1},x_{\gamma}\right)$, $n_{\gamma} = 2\zeta\left( 3\right)T^3/\pi^2$ and $\zeta\left( 3\right) = 1.202$. In order to produce a positive baryon to photon ratio, $\text{Im}\Big\{ \text{tr}\left[ m^{0}_{\nu}m^{*}_{\nu} \right]\Big\}$ should be negative.  It is worth noting the baryon asymmetry is dependent upon three quantities: the self-energy correction to the lepton propagator  represented by the loop factor $F\left( x_{1},x_{\gamma}\right)$, the effective neutrino mass matrices ($m_{\nu}$ and $m^{0}_{\nu}$) and finally the temperature, $T$, of the phase transition.  First, the loop factor is shown as a function of the  temperature normalised lepton energy ($x_{1}$) as shown in \figref{fig:loopfactor}. For Standard Model values of the  temperature normalised lepton thermal width, $x_{\gamma}\sim 0.1$, the loop factor provides an $\mathcal{O}\left(10 \right)$  enhancement to the lepton asymmetry. Second, the lepton asymmetry is crucially reliant on the effective neutrino masses. The structure of $m^{0}_{\nu}$ is determined by the particular high-scale flavour symmetry  and $m_{\nu}$ is the neutrino mass matrix which is diagonalised by the PMNS matrix $U_{PMNS}^{T}m_{\nu}U_{PMNS}=\text{diag}\left(m_{1}, m_{2}, m_{3} \right)$. This establishes a connection between the lepton asymmetry, low-energy neutrino parameters and the flavour symmetry. Finally,  in order to estimate the phase transition temperature, we assume $\text{Im}\{ \text{tr}[ m^{0}_{\nu}m_{\nu}^* ]\} $ is of the same order as $m_{\nu}^2\sim \left(0.1\text{eV} \right)^2$ and we have calculated that $F\left( x_{\gamma} \right)\sim \mathcal{O} \left(100 \right)$, hence the temperature for successful leptogenesis is 
\startcompact{small}
\begin{equation*}
T \sim 3 \sqrt{\eta_{B}}\, \frac{v_H^2}{m_\nu} \,.
\end{equation*}
\stopcompact{small}
In order to produce the observed baryon to photon ratio ($\eta_B=(6.19\pm0.15) \times10^{-10}$ \cite{Agashe:2014kda}) the temperature of the phase transition is approximately $T\sim 10^{11}$ GeV. The energy scale of this mechanism is similar to that of high-scale thermal leptogenesis. However, there are several improvements to this calculation which may lower the temperature.  These improvements include accounting for the differing charged lepton thermal widths and  calculating the fully time evolved asymmetry.

\section{Conclusion}\label{sec:conclusion}
CPPT is a completely new and novel mechanism that could simultaneously explain  the observed baryon asymmetry and  the  pattern of mixing in the lepton sector. Unlike conventional scenarios of leptogenesis, which specify a particular neutrino mass generation mechanism, CPPT allows for relative model independence as the new physics responsible for neutrino masses has already been integrated out before the CP-violating phase transition occurs. There are several interesting aspects of the  our mechanism that could be further explored. These include studying this mechanism in the context of a particular flavour model and  predicting the associated  gravitational wave spectra.

\end{document}